\documentclass{optica-article}
\usepackage{amsmath}
\usepackage{mathrsfs}
\usepackage[mathscr]{euscript}
\usepackage{gensymb}

\journal{opticajournal} 
\articletype{Research Article}
\usepackage{lineno}
\begin{document}
\title{Multifold enhancement of quantum SNR by using an EMCCD as a photon number resolving device}

\author{Rounak Chatterjee, Vikas S Bhat, Kiran Bajar, Sushil Mujumdar\authormark{*}}

\address{Nano-Optics and Mesoscopic Optics Laboratory, Tata Institute of Fundamental Research, 1, Homi Bhabha Road, Mumbai, 400 005, India}

\email{ \authormark{$*$}mujumdar@tifr.res.in} 

\begin{abstract*} 

The Electron Multiplying Charge Coupled Devices (EMCCD), owing to their high quantum efficiency and spatial resolution, are widely used to study typical quantum optical phenomena and related applications. Researchers have already developed a procedure that enables one to statistically determine whether a pixel detects a single photon, based on whether its output is higher or lower than the estimated noise level. However, these techniques are feasible at extremely low photon numbers ( $\approx0.15$ mean number of photons per pixel per exposure), allowing for at most one photon per pixel. This limitation necessitates a very large number of frames required for any study. In this work, we present a method to estimate the mean rate of photons per pixel per frame for arbitrary exposure time. Subsequently, we make a statistical estimate of the number of photons ($\geq$ 1) incident on each pixel. This allows us to effectively utilize the EMCCD as a photon number resolving device. This immediately augments the acceptable light levels in the experiments, leading to  significant reduction in the required experimentation time. As evidence of our approach, we quantify contrast in quantum correlation exhibited by a pair of spatially entangled photons generated by Spontaneous Parametric Down Conversion process. In comparison to conventional methods, our method realizes an enhancement in the signal to noise ratio by about a factor of 3 for half the data collection time. This SNR can be easily enhanced by minor modifications in experimental parameters such as exposure time etc. \end{abstract*}
\section{Introduction}
\label{Intro}
In recent decades, quantum optical phenomena have offered several high-utility working tools for technological purposes. For instance, effects such as single photon generation (including heralded single photon sources and quantum dots),  the Hong-Ou-Mandel effect \cite{HOM}, and  photon-pair entanglement in various bases like polarization \cite{Alain_aspect_PE, Kwait_PE}, angular momentum  \cite{OAM_E, OAM_E_rev}, and position-momentum  \cite{Howell_SE, Howell_SE2}, have proven to be of significant utility for research in areas like quantum information processing \cite{Qunatum-info-processing}. Additionally, these advancements have found applications in technological realms such as quantum cryptography  \cite{QKD_review, QKD_OAM_review}, quantum computing  \cite{QCom}. In the domain of fundamental research, one of the areas that utilizes the power of quantum concepts involves the study of disordered media using quantum light, in particular, entangled photon pairs  \cite{2P_speckle, 2P_complex_media_rev}. This research field includes the correction and certification of entangled states as they propagate through complex media \cite{SE_correction, SE_correction_rev}, followed by the implementation of quantum imaging  \cite{Q-imaging_rev}, leading to enhanced capabilities like super-resolution \cite{super_res}.

\par To effectively measure and harness the spatial entanglement of photons, the key requirement is of high quantum efficiency detectors with high spatial resolution. Typically, this process relies on \textit{raster scanning} of single photon avalanche diodes (SPADs) for spatial and temporal  detection \cite{Howell_SE, 2P_speckle}. Despite offering good quantum efficiency and excellent noise tolerance, this method is extremely time-consuming especially to get a good spatial resolution. A most desirable technique required in these quantum experiments is to develop a robust mechanism of data acquisition that can significantly reduce experimentation time. Existing photon number resolving  detectors can increase the quantum SNR \cite{QuantumSNR}, but they are limited in spatial resolutions. A standout instrument in the realm of low-intensity photon detection is the Electron Multiplying Charge Coupled Device (EMCCD), boasting high quantum efficiencies and high spatial resolution \cite{EMCCD_Description}. The device lacks temporal resolution, but this drawback is inconsequential for spatial entanglement measurements. Indeed, researchers have demonstrated the observation of spatial entanglement using EMCCDs \cite{EMCCD_spdc} and have also employed them for quantum imaging applications \cite{Q_img_EMCCD,Q_img_EMCCD2,Q_img_EMCCD3,Q_img_EMCCD4,Q_img_EMCCD5,Q_img_EMCCD6}. But due to lack of known photon number resolving techniques, they are usually used with very low light levels. On a similar note, recent technological progress has led to the development of SPAD cameras, combining the spatial resolution comparable to an EMCCD simultaneously with the temporal resolution of a SPAD. Effectively, this realizes the aspiration of a high-fidelity, high spatial resolution, quantum-sensitive camera\cite{SPAD_cam3,Hugo_SPADcam_1st} which is not limited to very low photon numbers per pixel. Although recent experiments have showcased their utility \cite{Hugo_SPADcam_exp,Padgett_SPADcam,SPADcam_measurements,SPADCam_LIIDAR}, this technology is still in its early stages and will depend on factors such as pixel size, pixel count, and overall cost-to-reward ratio.Apart from this, efforts have been made in realizing specialized equipment called Quanta Image Sensors that offer high quantum efficiency coupled with speed of CMOS technology.\cite{QIS1,QIS2}. These have been demonstrated to be sensitive to few-photon levels, however, these technologies are in the development phase and rather specialized.
\par Therefore, there is a growing need to develop alternative cost-effective techniques for a quantum imaging experiment to handle extremely low light levels, typically  $\leq 0.15$ photons per pixel per frame \cite{Low_pno_justification,Low_pno_justification2}. This imposes a lower bound on acquisition times. Even for sensitive devices such as EMCCDs, several hours and millions of frames are necessary to draw conclusive results. The primary reasons for these limitations can be attributed to two factors. The first is EMCCD's inability to intrinsically resolve multi-photon events. This entails the need to identify multi-photon pixels at a fair confidence-level, an ability that is hitherto unreported in current conventional analysis. Some attempts have been made in this direction\cite{EMCCD_physics} wherein the photoelectron number was fixed, and the EMCCD count distribution was derived therefrom. In our approach, we consider the EMCCD count value at par, and then derive the photoelectron distribution. Since EMCCD count is the actual measurand in any experiment, our method is more closely integrated with the experimental procedures. Secondly, systematic analysis is required to account for classical correlation effects that mask the quantum effects due to their prominence. Previously, researchers have studied the effects of multiphoton pixels in the specific case of momentum plane biphoton coincidences \cite{EMCCD_multiphoton_corr}. Our study is aimed at developing a generic strategy to identify multi-photon pixels, thus converting the EMCCD to an array-based photon number resolving device. We demonstrate that, with a proper classical correction strategy \cite{Photon_Corr_paper} in place, the recognition of multi-photon pixels can significantly enhance measurement contrasts and considerably reduce acquisition time. Furthermore, a photon number resolving camera is expected to allow for measuring various quantum enhanced protocols enabling new possibilities\cite{Multi-photon-corr}. Our multi-photon recognition scheme is based on systematic estimation of mean photon number per pixel per frame using a model to fit the EMCCD data. To that end, we employ a Bayesian-inferred probability function that identifies the thresholds to find incident photon numbers at a pixel. To demonstrate the usefulness of this distinction scheme, we perform a basic quantum correlation experiment \cite{EMCCD_spdc}. We observe that the results show a clear distinction in multi-photon events at arbitrary exposures. Using our multi-photon thresholding scheme, we show an enhanced contrast using a significantly lower number of acquired frames. Overall, we find that even a basic $2-$photon distinction almost doubles the contrast in the quantum correlation measurement for the same number of frames. With increasing photon number distinction, this contrast further increases till saturation is induced by the mean photon per pixel per frame. This significantly lowers experimental time and provides a major advantage to quantum optical experiments, given their ultra-sensitive nature to mechanical perturbations.

\section{The Mean photon number estimation model}
The EMCCD camera features a highly sensitive photon detector array (quantum efficiency $\geq 95\%$, pixel sizes $\mathcal{O}(10\mu$m)). The detected photoelectrons are transported via parallel and horizontal registers with adjustable clocks speeds to a gain register. The gain register consists of bits of cooled solid-state avalanche diodes with a graded conduction band that allows for \textit{only} impact ionization of electrons\cite{Matsuo1985NoisePA}. Ideally, each bit should double the incoming electron flux. However, in practice, the probability of duplication is deliberately kept small to ensure the stability of the avalanche process. This is followed by a pre-amplifier and an Analog-to-Digital(A/D) Converter that adds uniform bias to pixels and outputs positive count values \cite{EMCCD_Description}. We use an EMCCD with $512 \times 512$ pixels' frame transfer sensor array, with  a 552-bit gain register with adjustable electron multiplying gain.

\par To provide context for later sections, we  summarize the known physics of EMCCD cameras \cite{EMCCD_physics}. If the  stochastic duplication probability of a photo-electron is $p_c$, then for a gain register with $N_r$  bits,  the average gain$(G)$  is  $(1+p_c)^{N_r}$. For $k$ photoelectrons, the random number $X_k$ representing the final electron count has a probability distribution  
\begin{equation}\label{amplifier eqn}
    P_{X_k}(x|k)= \begin{cases}\delta(x),& k=0\\ \begin{cases} 0,&  x\leq 0\\ \frac{x^{k-1}e^{-\frac{x}{G}}}{G^k(k-1)!}, &x>0 \end{cases},& k\geq 1 \end{cases}
\end{equation}
where $\delta(x)$ represents the Kronecker$-\delta$ function, invoked to describe the situation wherein no additional electron is generated in absence of input photo-electrons. It is reasonable to assume that, for  a single exposure of the EMCCD frame, any given pixel $i$, upon illumination by photons, generate $k$ photoelectrons with a probability obeying Poisson statistics with a mean $\mu_{i}$. The amplified random count $X_{\mu_{i}}$ has  a distribution:
    \begin{equation}\label{eqn:Photon distributed}
    P_{X_{\mu_{i}}}(x|\mu_{i}) = \sum_{k=0}^{\infty}P_{X_k}(x|k)\mu_{i}^k \frac{e^{-\mu_i}}{k!}
    \end{equation}
Interestingly, equation \ref{eqn:Photon distributed} is valid for any photo-electron distribution. Although this model is an approximation, it suffices to describe an EMCCD with an ideal A/D converter, ideal pre-amplifier, and noiseless registers. The model can be fitted to sampled data after suitable subtraction of positive uniform offset provided by the A/D converter. However, in reality, EMCCD's suffer from dark noise, which has to be accounted for in a model, as discussed in the following subsection. 
\subsection{The Noise model}
\label{sec:Noise}
\begin{figure}[htb!]
\centering\includegraphics[width=0.7\linewidth]{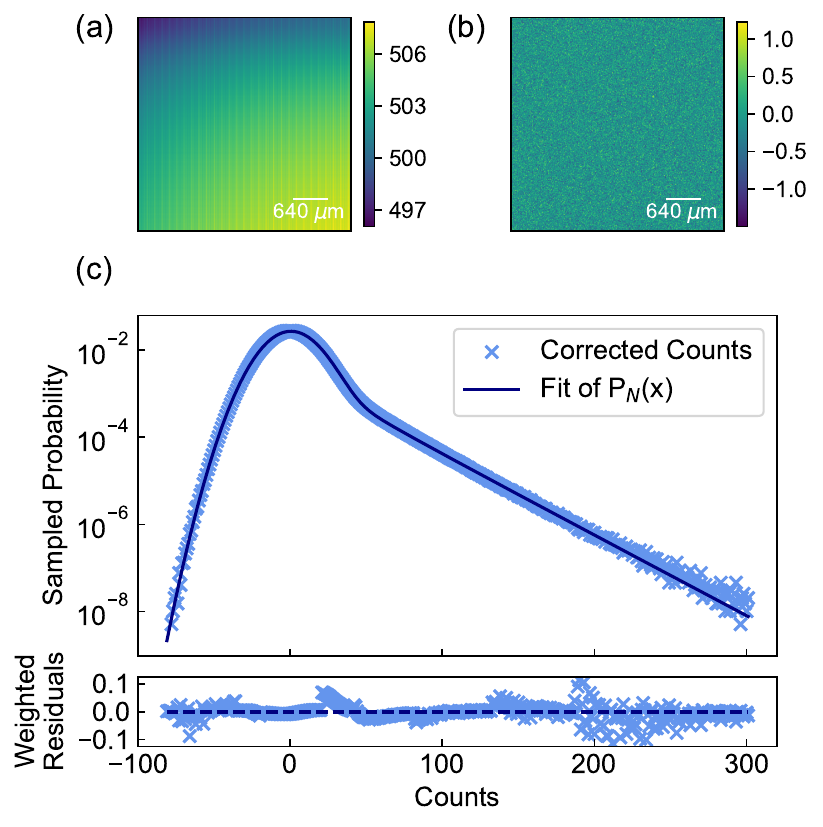}
\caption{Fitting of the Noise Model $(P_N(x))$  to the corrected noise data.(a)The correction image $(\bar{c}_{ij})$ estimated using the method in \cite{Modelling_inspiration_paper} (b) Mean image of corrected dark noise frames (c) Sampled Probability distribution of the Noise counts along with the model fit and corresponding weighted residuals.} 
\label{fig:Noise fit}
\end{figure}

In the conventional dark noise model\cite{Noise_Model_paper}, the amplification process is ideal with noise originating from spurious electrons. Apart from that, there also exists systematic noise attributed to the A/D converter, which introduces pixel-wise bias instead of a uniform offset\cite{Modelling_inspiration_paper}. To account for this bias, each pixel is assumed to provide an independent offset as computed by:
\begin{equation}
    \label{Correction_equation}
    b_{ij} = \langle \bar{r}_i + \bar{c}_j - m \rangle_{\mathrm{frames}}
\end{equation}
 where $b_{ij}$ is bias (dubbed as correction image)  , $\bar{r}_i$ is  $i^{th}$column average, $\bar{c}_j$ is $j^{th}$ row average,  $m$ is the average of the frame and $\langle\ldots \rangle_{\mathrm{frames}}$ is average over all frames. The correction image as seen in Fig~\ref{fig:Noise fit}a is constructed by acquiring 5000 images with shutter closed at particular EMCCD settings and then applying the above described procedure. This image is subtracted from every image be it noise or acquired photons. We can see from figure \ref{fig:Noise fit}b our correction image  accounts for this bias which is missed by methods used in \cite{Noise_Model_paper,Model_with_Biasing}.  The features in the correction image (systematic noise) arises from parallel and horizontal shift (clock) settings.         

\par As per conventional noise models \cite{Noise_Model_paper,Model_with_Biasing}, the dark noise is composed of three components. The first one are the spurious electrons that are generated during parallel transfer of photoelectrons, called clock induced charge (CIC), occurring with a small probability $p_{CIC}$. The second component are the spurious electrons arising in one of the bits of gain register with probability  $p_\mathrm{{ser}}$. Lastly, the third component comprises the normally distributed $(\sigma_{\mathrm{N}})$  pre-amplifer noise. Their distributions are as follows 
\begin{equation}
\begin{array}{cc}
     P_{X_{\mathrm{CIC}}}(x) = \begin{cases}
        0,&x<0\\
        1-p_{\mathrm{CIC}},&x =  0 \\
        p_{\mathrm{CIC}}\frac{e^{-\frac{x}{G}}}{G},&x>0
    \end{cases} &  
    ~~~~~~
     P_{X_{\mathrm{ser}}}(x) = \begin{cases}
        0,&x<0\\
        1-N_{r}p_{\mathrm{ser}},&x =  0\\
        p_{\mathrm{ser}}\sum_{m=1}^{N_{r}}\frac{e^{-\frac{x}{(1+p_{\mathrm{c}})^{N_{r}-m}}}}{(1+p_\mathrm{c})^{N_{r}-m}},&x>0
    \end{cases}
    \end{array}
\end{equation}

\begin{equation*}
    P_{X_{\mathrm{amp}}}(x) = \frac{1}{\sigma_\mathrm{N}\sqrt{2\pi}}e^{-\frac{x^2}{\sigma_\mathrm{N}^2}}
\end{equation*}

The final amplified count is given by the sum of all the random noise counts $X = X_{\mathrm{CIC}} + X_{\mathrm{ser}} + X_{\mathrm{amp}}$ and hence the fitting distribution is the convolution of the above three distributions.  We used a non-linear (Levenberg–Marquardt algorithm) least-squares fitting method to fit the resultant distribution to the experimental data.  Since count probabilities vary over a few orders of magnitude in the range, we used a weighted data fitting scheme wherein each count was weighed by their order of magnitude of probability\cite{Weighted_distribution_paper}.  For the particular data shown (blue crosses) in Figure \ref{fig:Noise fit}(c), a total of $5000$ image frames were captured with the shutter closed, using an exposure time of $15$ ms, a vertical shift time of $1.7 \mu$s, a horizontal shift rate of $10$~MHz, and a specified EM gain of $300$. The fit is shown by the black line. The goodness of the fit can be inferred from the weighted residuals, bound within $\pm0.1$, indicating an excellent fit. The obtained parameter values were $\sigma_N = 14.19$, $p_{CIC} = 0.0477$, $p_{ser} = 1.6 \times 10^{-4}$, and $p_c = 0.00573$.  The deviation of the specified gain value of $300$ from that estimated by $(1+p_c)^{N_r}$ ($\approx 100$) is attributed to a deviant definition of gain in the EMCCD specifications. \cite{Modelling_inspiration_paper,Gain_isnt_what_it_looks_paper}. This procedure provides us with the correction image and noise parameters.

\subsection{Incorporating the photons}
\label{sec:photons}
    
\begin{figure}[htb!]     

    \centering\includegraphics[width=1\linewidth]{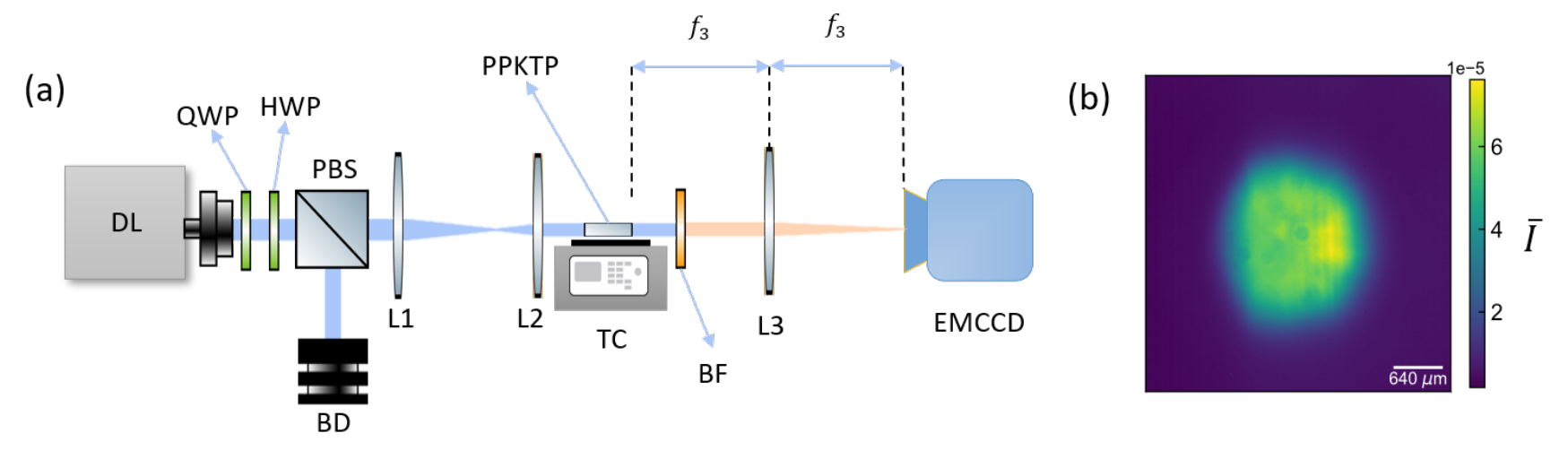}
    \caption{(a) The setup for measuring quantum correlation obtained from biphotons created from a nonlinear crystal. Emission from a  diode laser (DL) (wavelength $405\pm 0.1$~nm) is polarized by a combination of a quarter waveplate (QWP), a half waveplate (HWP) and a polarizing beamsplitter (PBS), and then demagnified  with lenses L1 and L2. It is injected into a periodically-poled  potassium titanyl phosphate crystal (ppKTP,  $10$mm $\times$ $2$mm $\times$ $1$mm) maintained at $40\degree$C  by a temperature controller (TC) to emit degenerate biphotons at  wavelength of $810$~nm. The bandpass filter (BF, $810 \pm 10$nm) blocks the pump beam and  the biphotons are led to the EMCCD using a lens (L3) that maps the biphotons' probability density function in momentum variables.  (b) The normalized spatial intensity distribution obtained from the EMCCD.}
    \label{fig:Setup}
    \end{figure}

Next, we acquire photons  opening the EMCCD shutter(refer figure \ref{fig:Setup}) using the same settings as for noise. The source generates spatially entangled photons which show quantum correlation (see section \ref{sec:Momentum Correlation} for details) . For descriptions in the following paragraphs, the data can be acquired anywhere after the crystal. A typical data set consists of  choosing $k \times l$ pixels on the EMCCD as a relevant area of interest and then acquiring $M$ number of  frames and corrected for systematic noise (prescribed in \ref{sec:Noise}) to give us $X_j (j = 1~ \text{to}~M)$.

\begin{figure}[htb!]        
            \centering\includegraphics[width=0.9\linewidth]{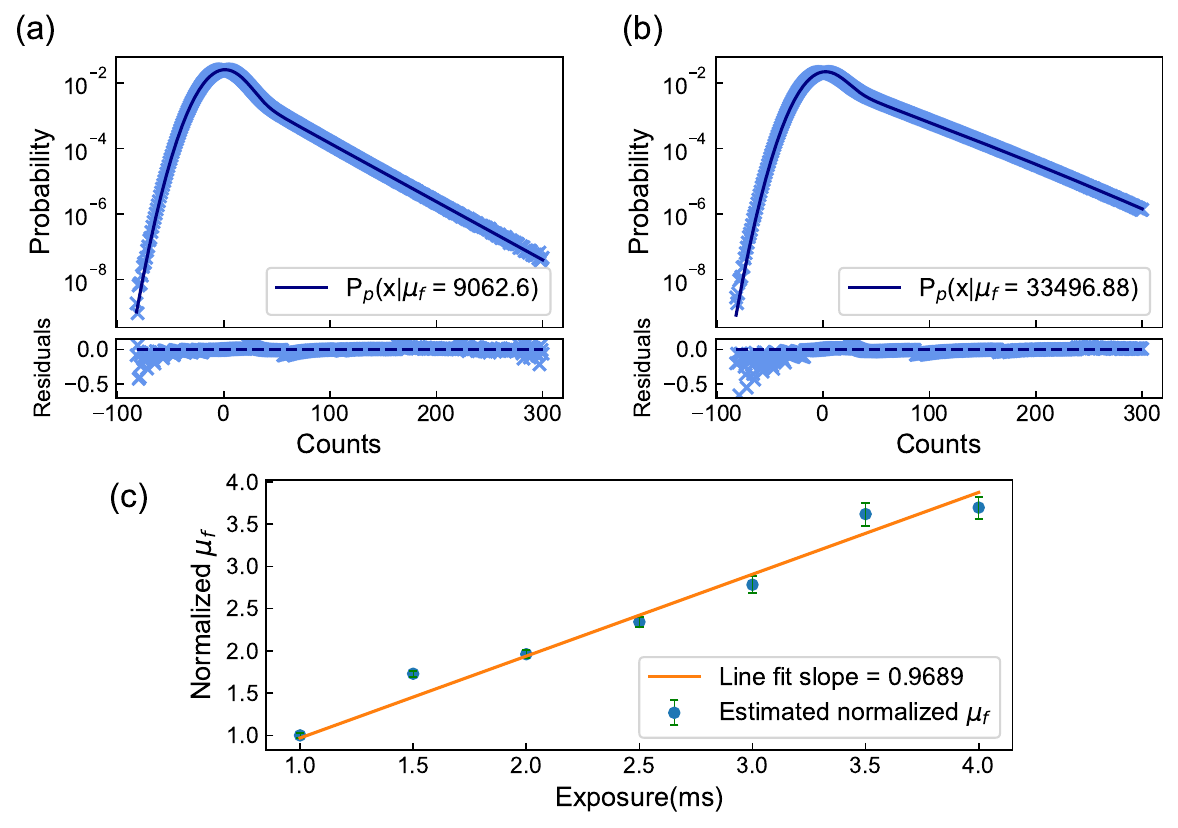}
                \caption{Fitting of the Photon Model $(P_p(x))$  to the corrected noise data.(a) Sampled distribution(blue crosses) and model fit(black solid) along with variation of residuals shown for estimated lowest mean photon number. (b) Similar depiction as (a) but for highest estimated mean photon number. (c)Variation of estimated mean photon number ($\mu_f$) with exposure time of EMCCD. The $\mu_f$ values are normalized with respect to the minimum value estimated at $1$ms exposure ($\approx 9062$).} 
            \label{fig:Photon fit}
        \end{figure}

The first step in estimating photons is constructing a normalized spatial distribution of photons on the area of  interest. To do this an averaged frame ($\sum X_j/M$)  is obtained, which is further divided by the sum over all pixels to provide a normalised spatial intensity distribution $\bar{I}$, depicted in \ref{fig:Setup}(b). For a  pixel $i$ with coordinates $(x_i, y_i)$, $\bar{I}(x_i, y_i)$ corresponds to the normalised intensity on that pixel. We proceed with the plausible assumption that, if the mean number of photons falling on an individual frame is $\mu_f$, then  the $i^{th}$ pixel sees Poisson-distributed photons with the mean $\mu_i = \mu_f\bar{I}(x_i, y_i)$. \ref{fig:Photon fit}(a) depicts, in blue crosses, the histogram of counts over all pixels and all corrected frames, i.e, over $k\times l \times M$ values.  \ref{fig:Photon fit}(b) represents the same, for a higher exposure time of the EMCCD.  The random count $X_i$ in each pixel is sum of noise counts $(X_N)$ and photon counts$(X_{\mu_i})$. Towards a numerical fit of these distributions, we employ the equation 
    \begin{equation}\label{eqn: Frame distributed photons}
    P_{p}(x|\mu_f) = \frac{1}{k\cdot l}\sum_{i =1}^{k\cdot l}P_N(x) \star P_{X_{\mu_i}}(x|\mu_i = \mu_f \bar{I}(x_i,y_i))
    \end{equation}
where $P_N(x)$ is the noise distribution and $P_{X_{\mu_i}}(x|\mu_i)$  is photon distribution   of equation \ref{eqn:Photon distributed}, and the $\star$ represents convolution. Since each pixel is independent, the final probability $P_p$ is the sum of individual probabilities.  
\par The above distribution is parameterized by  $\mu_f,p_\mathrm{c},\sigma_\mathrm{N}, p_{\mathrm{CIC}}$ and $p_{\mathrm{ser}}$ with the first two parameters characterizing the signal photons. Using the facts that (i) $P_N(x)$ is independent of pixels and can be pulled out of the summation, and (ii) convolution is distributive, we can simplify equation \ref{eqn: Frame distributed photons} as \cite{Modelling_inspiration_paper}:
        \begin{equation}\label{eqn:Main fitting}
            P_P(x|\mu_f) = P_N(x) \star \frac{1}{k \cdot l}\sum^{k \cdot l}_{i=1}\begin{cases}
                e^{-\mu_i},&x = 0 \\
                e^{-\frac{x}{G}-\mu_i}\sqrt{\frac{\mu_i}{Gx}} I_{1}\left(2\sqrt{\frac{x\mu_i}{G}}\right),& x>0
            \end{cases}
        \end{equation}
Where  $I_1$ is the Modified Bessel function of the first kind, order 1. This model, called the Poisson-Gamma-Normal Model (\textit{PGN model})\cite{Modelling_inspiration_paper}, was introduced for single pixel  photon counting. We extend this work for multiple pixels by  considering  the spatial distribution $(\bar{I})$ of intensity over a frame. Thereby, equation \ref{eqn:Main fitting} becomes the main fitting equation for a general  spatial distribution of photons incident on the EMCCD.

\par We fix the parameters $\sigma_\mathrm{N}, p_\mathrm{{CIC}},$ and $p_\mathrm{{ser}}$ as estimated from noise  and keep $\mu_f$ and $p_c$ as free parameters for the photon distribution fitting. Figure \ref{fig:Photon fit}(a) and \ref{fig:Photon fit}(b) show the respective fits for exposures of $1$~ms and $4$~ms. Evidently,  the shape of the distribution changes with changing incident intensity, and the fitting routine succeeds in emulating the change, as seen from the solid black line. The excellence of the fits is evident from the residuals shown below. As a confirmatory test, we study the variation of $\mu_f$  with  exposure time (figure \ref{fig:Photon fit}(c)), and recover a linear variation as expected. In this study, we needed to maintain $p_c$ as a fitting parameter, despite having extracted it once in the noise fitting as discussed in \ref{sec:Noise}. This enabled us to obtain the excellent fits shown in the figures \ref{fig:Photon fit}(a) and \ref{fig:Photon fit}(b). We found that $p_\mathrm{c}$ decreases with increasing $\mu_f$. It can be speculated that the EMCCD gain behaves differently depending on the amount of light falling on the sensor.  Although the change in $p_\mathrm{c}$ is  only around $2.3\%$, it suggests  some probable mechanism that tweaks the gain to  ensure stability of the amplification process to keep EMCCD safe and long-lived \cite{EMCCD_Description}.

\section{Thresholding method}
\label{sec:thresholding}
With the fit parameters calculated in the previous section, we can now calculate the \textit{probability of a photoelectron number $k$} being observed given a count $x$ at pixel $i$. Since we have the analytical form of probability of EMCCD counts$(x)$ at the $i^{th}$  pixel given as $P_i(x|k,\mu_i) = P_N(x) \star P(x|k)P(k|\mu_i)$, we can employ Bayes' theorem\cite{Bayes_Answer} to pose the above question mathematically as:
\begin{equation}
\label{eqn: Bayes inference}
    P_i(k|x,\mu_i) = \frac{P_N(x) \star P(x|k)P(k|\mu_i)}{\sum_{k=0}^{\infty}P_N(x) \star P(x|k)P(k|\mu_i)} = \frac{P_N(x) \star P_{X_k}(x|k)\frac{e^{-\mu_i}\mu_i^k}{k!}}{P_N(x)\star P_{X_{\mu_i}}(x|\mu_i)}
\end{equation}
This is obtained using equations \ref{amplifier eqn} and \ref{eqn:Photon distributed}. The denominator was simplified like equation \ref{eqn:Main fitting} for the concerned pixel $i$. We can see from equation \ref{eqn: Bayes inference} that noise is coupled in the Bayesian expression. Nonetheless, the noise only serves as a baseline accounting for the spurious counts. 

\begin{figure}[htb!]
    \centering\includegraphics[width=0.9\linewidth]
    {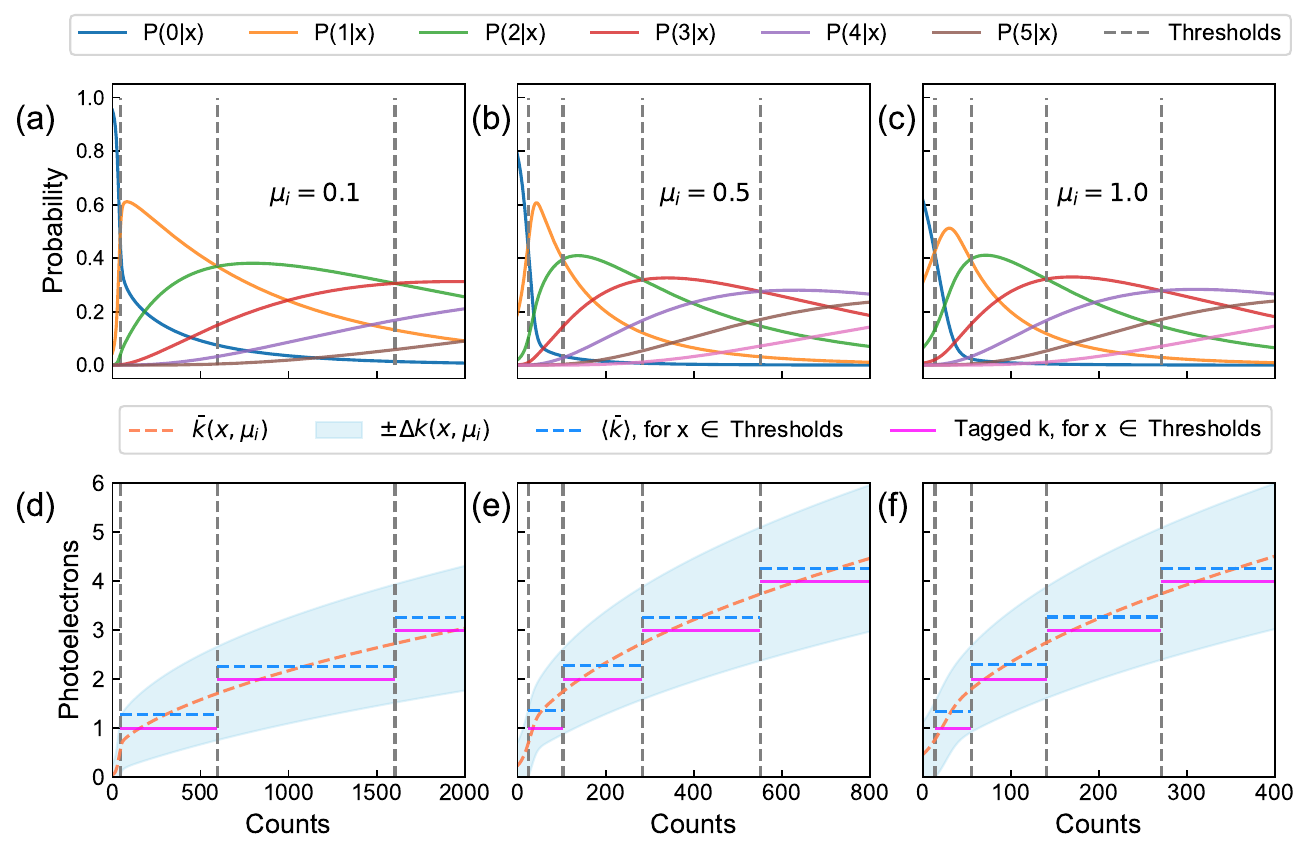}
    \caption{Bayesian-inferred function of  photon number probability (follow legend for colours of functions) on a pixel given counts$(x)$ and mean photon number $(\mu_i)$ as given by equation \ref{eqn: Bayes inference} . They are plotted as function of  counts for three different fixed values $\mu_i = $ (a) $0.1$ (b) $0.5$  (c) $1.0$ . The grey lines show various computed threshold values for photon numbers.The lower panel shows the expected mean photoelectron number $\bar{k}$ (dotted orange) along with the standard deviation  ($\bar{k}\pm \Delta k$ blue shaded region) for three different $\mu_i$ = (d) $0.1$ (e) $0.5$ (f) $1.0$. Within each threshold region the mean photon number (dashed blue) and tagged photoelectron number (magenta) are also shown.}
    \label{fig:Threshold}
    \end{figure}
\par The plots of the Bayesian-inferred functions in equation \ref{eqn: Bayes inference} for different $\mu_i$ (noise characteristics and gain taken from analysis in section \ref{sec:photons}) are shown in Figures \ref{fig:Threshold}(a), (b), and (c). Each plot corresponds to a different $\mu_i$, as labeled. The vertical dotted lines depict the threshold points wherein a certain photon number dominates. The left and right threshold points for the  $k^{\mathrm{th}}$ photon number is where the probability of $k$ is the same as that of  $k-1$ and $k+1$ respectively. \[P_i(k-1|x = T_k^{left}) = P_i(k|x = T_k^{left}) \]\[P_i(k|x = T_k^{right}) = P_i(k+1|x = T_k^{right})\]
where $T_k^\text{left}$ and $T_k^\text{right}$ represent the threshold counts for photon number $k$. For the zero photoelectron count, the lower threshold is always taken as zero. For instance, in figure \ref{fig:Threshold}(a), the point at which the green curve overtakes the orange curve at count of approximately $550$ is assumed as the threshold between photon count $2$ and $1$. Essentially, this implies that when measured counts are in the range $\approx550$ to $\approx1550$, they are most likely to have arisen from $2$ incident photons. Clearly, these ranges vary with $\mu_i$. To justify the choice of thresholds for a given photoelectron number,  we compute, using equation \ref{eqn: Bayes inference}, the mean photoelectron number $\bar{k} = \sum_{k}{k P_i(k|x,\mu_i)}$ and the standard deviation $\Delta k = \sqrt{\sum_{k}{(k-\bar{k})^2P_i(k|x,\mu_i)}}$ as a function of observed counts$(x)$ and mean photon per pixel $(\mu_i)$. Using the distributive property of convolution, one can conclude  from equation \ref{eqn: Bayes inference} that the sum over $k$ can be done only on $P_{X_k}(x|k)\frac{\mu_i^k}{k!}$ , giving rise to following expressions for $\bar{k}$ and $\bar{k^2}$ as:
\begin{equation*}
    \bar{k}(x,\mu_i) = \frac{P_N(x) \star e^{\left(-\frac{x}{G}-\mu_i\right)}\frac{\mu_i}{G}I_0\left(2\sqrt{\frac{x\mu_i}{G}} \right)}{P_N(x)\star P_{X_{\mu_i}}(x|\mu_i)},
\end{equation*}
\begin{equation}
    \bar{k^2} = \frac{P_N(x) \star e^{\left(-\frac{x}{G}-\mu_i\right)}\frac{\mu_i}{G^{3/2}}\left(\sqrt{G}I_0\left(2\sqrt{\frac{x\mu_i}{G}} \right) + \sqrt{x\mu_i}I_1\left(2\sqrt{\frac{x\mu_i}{G}} \right)\right)}{P_N(x)\star P_{X_{\mu_i}}(x|\mu_i)},
\end{equation}
where $I_0$ and $I_1$ are modified Bessel functions of first kind of order zero and one respectively. The plot of  $\bar{k} \pm \Delta k$ as a function of counts$(x)$ is shown in figure \ref{fig:Threshold}(d),(e) and (f) for various $\mu_i$ values. To get an estimate of mean number within the thresholds, that is $x \in [T_k^{left},T_k^{right}]$ , we notice that for each $x$ and $\mu_i$ the distribution is normalized in $k$ $(\sum_k{P_i(k|x,\mu_i}) = 1)$. Hence for an arbitrary (unbiased) choice of $x$ observed experimentally, we can define a normalized distribution in the threshold region as $\frac{1}{(T_k^{right} - T_k^{left})}\sum_x{P_i(k|x,\mu_i})$. Thus the mean photoelectron number within the threshold becomes a simple average given as $\langle \bar{k}(\mu_i,T_k^{left},T_k^{right})\rangle =\frac{1}{(T_k^{right} - T_k^{left})}\sum_x{\bar{k}(x,\mu_i)}$.  It can be seen clearly from figure  \ref{fig:Threshold}(d),(e) and (f) that the expected value within the threshold (blue dashed line) lies very close to the tagged photoelectron value (magenta line). The standard deviation (blue bands) lies  between 1 and 2, which comes from the  overlap of the photon number distributions. The architecture of an EMCCD doesn't allow for decoupling of the distributions unlike in devices like the QIS\cite{QIS1} or the SPAD camera\cite{Hugo_SPADcam_1st}. Nonetheless, the distributions are peaked distinctly  enough to extract information about the photon numbers.  Thus, based on a maximum likelihood estimate, our method provides an optimal way to extract photon numbers from EMCCD counts, which is validated by the enhancement in quantum contrast measurements discussed in the section \ref{sec:Momentum Correlation}.

We found that a numerical computation of $\langle \bar{k} \rangle$ for  various values of gain at a fixed $\mu_i$ show that with increasing gain the expectation draws closer to the tagged value, suggesting that higher gain value can improve the accuracy of estimation. Here, we encounter a trade-off: higher gain also implies that the higher \textit{k} values are pushed to larger EMCCD counts. However, in a practical scenario, it is not advised to record very high counts per pixel on the EMCCD on grounds of integrity of the sensor. Therefore, there exists a region of compromise wherein the gain is maintained moderate enough. In these experiments, we avoided working close to saturation value of the EMCCD ($2^{16}-1 = 65535$). Nonetheless, we reached a photon number resolution up to about 16 which corresponds to a count of about 7000 at a gain of 300. 

In our scheme, we consider only those photoelectron numbers whose Poisson probability is greater than $5\%$ for given $\mu_i$. While this limit is somewhat arbitrary, it is justified because all the main physical effects are captured by the photon numbers around the mean number. 

\begin{figure}[htb!]
    \centering\includegraphics[width=0.9\linewidth]
    {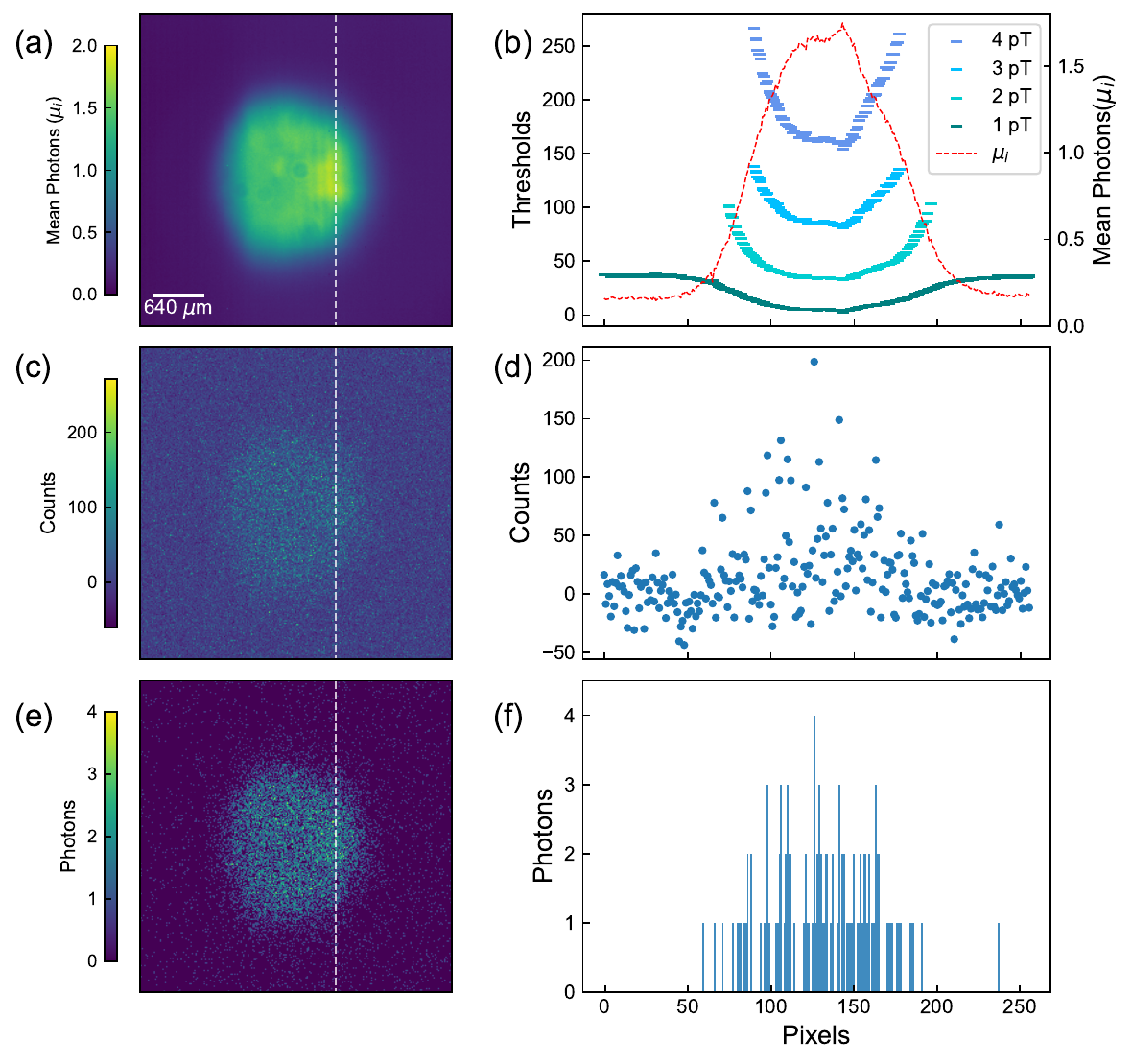}
    \caption{Creation and use of \textit{Threshold map}. (a) Mean photon number at all pixels of the EMCCD camera. (exposure $3$~ms)   (b) Mean photon number ($\mu_i$, red dashed)  for  pixels along the white dashed lines in (a). The various coloured dashes (color coding given in the legend) show the threshold points at each pixel. (c) Corrected EMCCD counts from an arbitrarily chosen frame. (d) Cross-section of fig (c) along the white dashed line, same as in (a). (e) Estimated photon number on each pixel for the frame shown in (c). (f) Cross-section of fig (e) along the white dashed line computed using threshold map (b) and corrected EMCCD counts in (d).}  
    \label{fig:Threshold2}
    \end{figure}

\par Next, we utilize this system of  photon thresholding to  identify the number of photons falling onto the various pixels.  Using the normalized spatial distribution of intensity $(\bar{I})$ as described in section \ref{sec:photons} and a $\mu_f$ for a certain exposure, we can create a pixel map for mean photons, i.e $\mu_i(x_i,y_i) = \mu_f \bar{I}(x_i,y_i)$. Image \ref{fig:Threshold2}(a)  depicts this map for an exposure time of $3$ ms, with estimated  $\mu_f \approx 2.52 \times 10^{4}$.  The red dashed curve in subplot (b) is the cross-section along the white dashed line in (a), or essentially, the $\mu_i$ at the pixels marked by the white dashed line.  Next, using the procedure described above, we estimate the thresholds along the curve. For brevity of text, we create a labelling convention wherein a term   "$2pT$"  implies thresholding with number of photons $= 2$. The multiple coloured dashes shows the threshold values specified in the legend.   Image \ref{fig:Threshold2}(c) illustrates the corrected EMCCD counts from an arbitrarily chosen frame, while plot \ref{fig:Threshold2}(d) shows the cross-section of (c) along the white dashed line, i.e, along the same pixels used above in thresholding.  Finally, image \ref{fig:Threshold2}(e) illustrates the counted photons for the frame (c), and (f) shows the same along the white dashed line.  This procedure enables us to \textit{probabilistically} identify the number of photons falling on pixels of the EMCCD, after knowing the count values.  A simple comparison between (c) and (e) reveals a significant increase in the contrast of  acquired data. Such photon-counted frames can be employed in different studies such as bi-photon correlation experiments, quantum interference effects, or the study of spatial entanglement, as well as related applications like quantum imaging, etc. 

\begin{figure}[htbp!]
    \centering\includegraphics[width=0.85\linewidth]{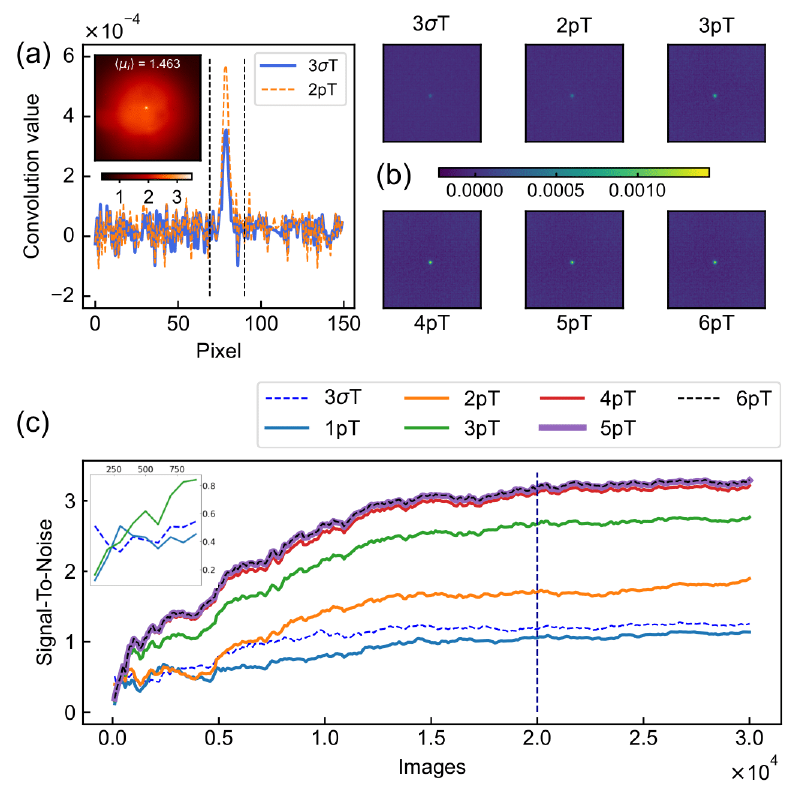}
\caption{The comparison of Signal-to-Noise ratio in quantum correlation experiment.(a) Comparison of momentum correlation shown by the SPDC beam in the inset, computed from  data thresholded using  \textbf{$3\sigma T$} method (blue solid line) and and our method (orange dashed lines). (b) Panels showing changing contrast in observed momentum correlation with addition of more threshold points and compared with the $3\sigma T$ method. The convolutions  are computed as mean over 30,000 frames. (c) Shows variation of signal-to-noise ratio with increasing observation frames for the $3\sigma T$ method (blue dashed)  and our method for various thresholds (follow  label for color coding). The inset shows zoomed in version for first 900 frames. The Signal-to-Noise ratio is computed by equation \ref{eqn:SNR}, about the two black lines in fig\ref{fig:Convolution}(a). The deep blue vertical line at $2 \times 10^4$ shows the enhancement of SNR for increasing threshold values.  }  
    \label{fig:Convolution}
    \end{figure}

We emphasize that, within the safe limits of the EMCCD, i.e, working relatively far from the saturation count, this technique provides a significant advantage. In our experiments wherein we achieved a photon number resolution up to about 16, we already realized a shortening in the experimentation time by a factor of about 3-4 compared to binary thresholding, and sometimes by factors of 5-6 in other experiments.

\section{Momentum correlations with new thresholding method}
\label{sec:Momentum Correlation}
To  demonstrate the capabilities of our technique, we consider a quantum correlation experiment\cite{EMCCD_spdc}. As described in Section \ref{sec:photons}, the source used for our experiments generates pairs of photons. These pairs are entangled spatially, with a wavefunction that can be approximated as a double Gaussian\cite{Gaussian_wf,Gaussian_wf2} that exhibits spatial and momentum correlation among pairs\cite{SPDC_review,SPDC_review2}.

\begin{equation}
\Psi(\mathbf{r}_1,\mathbf{r}_2) \propto e^{-\left(\frac{\mathbf{r}_1+\mathbf{r}_2}{\sqrt{2}\sigma_{+}}\right)^2} e^{-\left(\frac{\mathbf{r}_1-\mathbf{r}_2}{\sqrt{2}\sigma_{-}}\right)^2}
\end{equation}

where $\boldsymbol{r}_1,\boldsymbol{r}_2$ are transverse coordinates of the photons and $\sigma_+$ and $\sigma_-$ are related to  momentum and spatial correlations. Thus, a spatially-resolving single photon detector can reveal anti-correlations if  we image the Fourier plane of the crystal face, which measures the momentum variable. This is exactly done by putting the EMCCD at focal plane of lens $L3$ as shown in figure \ref{fig:Setup}(a). We acquired $30000$ frames with an $50$ms exposure time  and auto-convolved it (Mean photon distribution of  this beam shown in inset of figure \ref{fig:Convolution}(a)).

Since the photon pairs are correlated in time, their quantum correlation exists within the same image. To remove the classical correlations, the cross-convolution of the image with the subsequent one is subtracted from the auto-convolution.  To improve the signal-to-noise ratio, a binary thresholding method is employed, wherein all EMCCD counts below a the threshold are set to $0$ and those above are set to $1$\cite{EMCCD_Description}. This one-photon threshold point is set at a count value $n\sigma_N$ higher than the mean count value. In our case, we use $n = 3$. For brevity of text, we create a labelling convention wherein a term  "$3\sigma T$" implies $n\sigma$-thresholding with $n=3$, as compared to a term "$2pT$" which implies the above-mentioned photon-thresholding(section \ref{sec:thresholding})  with number of photons equal to $2$. The blue curve in the main plot displays (a cross-section along the central row of pixels of the) quantum correlation computed on this beam profile using the  $3\sigma T$ system. On the other hand, the dashed orange curve depicts the same as calculated from our $2pT$ system. A clear enhancement compared to the standard $3\sigma T$ is observed, as quantified by  signal-to-noise ratio  
    \begin{equation}
    SNR = \frac{\langle \mathrm{Signal}\rangle - \langle \mathrm{Noise floor} \rangle}{\mathrm{std(Noise floor)}}
    \label{eqn:SNR}
    \end{equation}

\par The signal is defined within the black dashed lines. In Figure \ref{fig:Convolution}(b), the multiple panels display the situations with various multi-photon thresholding as labeled. Clearly, our method outperforms conventional  $3 \sigma T$ model even with most basic $2pT$ scheme. In other words, the improvement in contrast implies that similar quantum correlations can now be drawn with fewer acquired images. Figure \ref{fig:Convolution}(c) summarises this scenario.  The figure compares the number of frames required to compute the SNR between various methods. Focusing on the vertical dashed line that identifies the situation at $20000$ images, the SNR is seen to improve by a few factors with $kpT$. The saturation in the curves implies an optimal number of frames needed to be grabbed. At the same time, the curve separations indicate an optimal thresholding that can be applied. For instance, the reward at $4pT$ and $5pT$ is very comparable, suggesting $4pT$ as an optimal threshold.

\section*{Conclusions}
In summary, we have proposed and verified a multi-thresholding algorithm in the photon counting process using an EMCCD. This can significantly enhance the efficiency of an EMCCD used in quantum experiments. Instead of solely studying noise properties of the EMCCD to enhance its robustness as a photon counter, our aim was to unveil the information hidden within its amplification system and leverage it to our advantage. We have developed a physically plausible system for experimental deduction, regardless of the photon number distribution in time or space. It is evident that all the analysis  that is  applicable to single-photon states can be extended to multi-photon events since the underlying physics remains unchanged. However, the most challenging aspect lies in compensating for higher-order classical effects. One major contribution provides a generalized model for measuring photon detection correlations and deriving bi-photon joint probability density \cite{Photon_Corr_paper}. We believe that our EMCCD multi-photon thresholding model can directly complement the already theorized analysis techniques. Our work demonstrates that not only the acquisition time is decreased but also the contrast is significantly increased when accounting for multi-photon events. Since quantum optical experiments, are rather sensitive to mechanical perturbations in optical setup due to vibrations, air currents etc, reducing the experimental time is a major desirable in this field. In particular, our work stands to  directly benefit the  field of quantum imaging, quantum information processing and quantum communications.

\par Undoubtedly, instruments like raster scanning SPAD systems and the and  SPAD cameras \cite{Hugo_SPADcam_1st} are good choices for quantum experiments, but they have their own limitations such as available pixel count, acquisition time, and especially, the cost of  resources required. In this context, EMCCD is currently at the peak of its demand, offering a reasonable balance in terms of sensitivity, pixel resolution, coverage, and acquisition time, all within a sustainable cost range. Our analysis ensures that the full potential of this robust device is realized. Our future plans for this work include incorporating effects like blooming into our analysis to improve recognition of photon states.

\section{Acknowledgements}
We acknowledge the constant support and inspiration given by all the members of Nano-Optics and Mesoscopic Optics Laboratory, Tata Institute of Fundamental Research, India. We also express our gratitude to the Department of Atomic Energy, Government of India, for funding for Project Identification No. RTI4002 under DAE OM No. 1303/1/2020/R\&D-II/DAE/5567,Ministry of Science and Technology, India. The authors declare no conflict of interest.

\section{Disclosure}
The authors declare that there are no conflicts of interest related to this article.

\section{Data Availability}
Data and related code for analysis for this procedure is available upon reasonable request.

\bibliography{References}

\end{document}